%% file: main.tex
\documentclass[
	a4paper, % Paper size, use either a4paper or letterpaper
	10pt, % Default font size, can also use 11pt or 12pt, although this is not recommended
	unnumberedsections, % Comment to enable section numbering
	twoside, % Two side traditional mode where headers and footers change between odd and even pages, comment this option to make them fixed
]{LTJournalArticle}

\runninghead{} % A shortened article title to appear in the running head, leave this command empty for no running head

\footertext{\textit{RISC-V Summit Europe, Paris, 12-15th May 2025}} % Text to appear in the footer, leave this command empty for no footer text

\usepackage[inkscapearea=page]{svg}
\usepackage{graphicx}% http://ctan.org/pkg/graphicx
\usepackage{hyperref}
\usepackage{amsmath,amssymb,amsfonts}
\usepackage{siunitx}
\usepackage{float}

\titlespacing{\section}{0pt}{*0.8}{*0.8}

\usepackage[toc, section=section]{glossaries}
\makeglossaries
\input{acronyms}

\title{Work-In-Progress: Accelerating Numpy \\ With OpenBLAS For Open-Source RISC-V Chips} % Article title, use manual lines breaks (\\) to beautify the layout

\if 11
\author{
	Cyril Koenig\textsuperscript{1}\thanks{Corresponding author: \href{mailto:cykoenig@iis.ee.ethz.ch}{\tt cykoenig@iis.ee.ethz.ch}}, Enrico Zelioli\textsuperscript{1}, Frank K. Gürkaynak\textsuperscript{1} and Luca Benini\textsuperscript{1,2}
}
\date{\footnotesize\textsuperscript{\textbf{1}}ETH Zurich\\ \textsuperscript{\textbf{2}}Università di Bologna}

\else

\author{%
	Authors omitted for blind review
}

\fi

\renewcommand{\maketitlehookd}{%
	\begin{abstract}
		\noindent RISC\nobreakdash-V allows for building general-purpose computing platforms with programmable accelerators around a single open-source ISA. However, leveraging heterogeneous SoCs within high-level applications is a tedious task. In this preliminary work, we modify the OpenBLAS library to offload selected linear kernels to a \gls{pmca} using OpenMP. By linking the Python package Numpy against this library, we enable acceleration of high-level applications. We target an open-source heterogeneous System-on-Chip with a \texttt{rv64g} Linux capable host and a \texttt{rv32imafd} \gls{pmca}. Using this platform emulated on FPGA, and the presented software stack, we can accelerate Phyton applications with linear algebra operators like matrix multiplication.
	\end{abstract}
}

\begin{document}

\maketitle % Output the title section

%----------------------------------------------------------------------------------------
%	ARTICLE CONTENTS
%----------------------------------------------------------------------------------------

\glsresetall
\glsunset{isa} \glsunset{cpu} \glsunset{gpu} \glsunset{soc} \glsunset{api} \glsunset{iommu} \glsunset{blas} \glsunset{dram} \glsunset{fpu} \glsunset{fpga} \glsunset{dma} \glsunset{simd}
\section{Introduction}
With the democratization of artificial intelligence and machine learning, the demand for embedded and high-performance hardware optimized for linear calculus is continuously growing. In this context, RISC\nobreakdash-V will allow for building general-purpose platforms with linear computing accelerators from different vendors around the same open \gls{isa}.

Nevertheless, accelerating applications written in high-level languages can be a tedious task. The \gls{blas} \gls{api} addresses this issue by identifying a list of basic linear algebra operations. Multiple implementations of the \gls{api} have been proposed, and many high-level applications can leverage their platform-specific optimizations by binding linear algebra operators (e.g., matrix multiplication).

These implementations can target different purposes. OpenBLAS, for instance, features hand-crafted kernels for various architectures and \glspl{isa}. BLIS, offers high performance for symmetric multi-threading\,\cite{blis_perf}. IRIS-BLAS\,\cite{iris_blas} targets heterogeneous architectures with general purpose \glspl{gpu}. However, there is no open source implementation combining hand-crafted RISC\nobreakdash-V host kernels and heterogeneous computation with RISC\nobreakdash-V \glspl{pmca}.

In this work, we use the Hero \gls{sdk} presented in \cite{herosdk} to extend OpenBLAS for an open-source RISC\nobreakdash-V based \gls{hesoc}. We add a heterogeneous implementation of \gls{gemm} for a \texttt{rv32imafd} \gls{pmca}. We benchmark our accelerated OpenBLAS implementation from a Python application running on the \texttt{rv64g} host-core. Our preliminary result show $2.71\times$ execution speedup when offloading a Numpy matrix multiplication on the device rather than executing it on the host.

%------------------------------------------------

\section{Open-Source Platform}

\begin{figure}[ht]
\centering
\includegraphics[width=0.95\columnwidth]{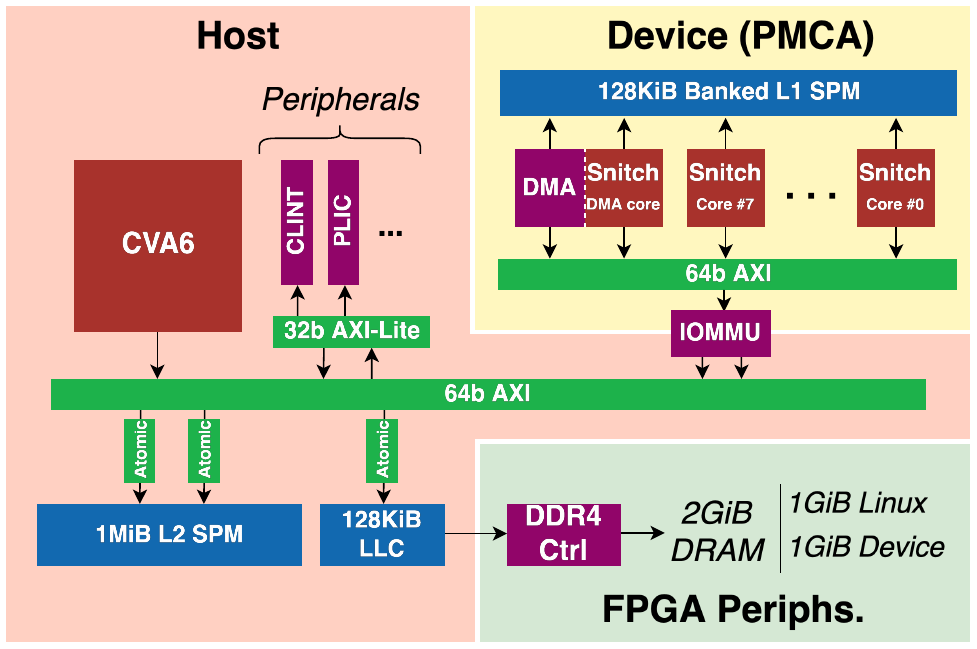}
%\includesvg[width=0.95\columnwidth]{fig/openblas-archi.drawio.svg}
\caption{Open-Source heterogeneous platform with CVA6 and Snitch. The L1 \acrshort{spm} contains the device local data, the dual-port L2 \acrshort{spm} contains constants and device instructions, and the device \acrshort{dram} contains physically contiguous buffers for shared data structures.}
\label{fig:platform}
\end{figure}

The platform\footnote[1]{\tiny{\href{https://github.com/pulp-platform/carfield/tree/date_iommu_evaluation}{https://github.com/pulp-platform/carfield/tree/date\_iommu\_evaluation}}} used in this work is based on Cheshire, a \gls{soc} built around the \texttt{rv64g} application-class core CVA6. This host is coupled to a \gls{pmca} that features eight Snitch cores with double precision \glspl{fpu}. The accelerator cluster contains \SI{128}{\kibi\byte} of local \gls{spm} that is refilled from the external shared \gls{dram} using a \gls{dma} engine. The \gls{dram} is partitioned into two regions, one used by the operating system, and one manually managed to avoid fragmentation. When the \gls{iommu} is not used, shared data structures must be copied to the device \gls{dram} before use. We emulate the platform on a Xilinx VCU128.

\begin{figure*}[ht]
    \centering
    \includegraphics[width=\textwidth]{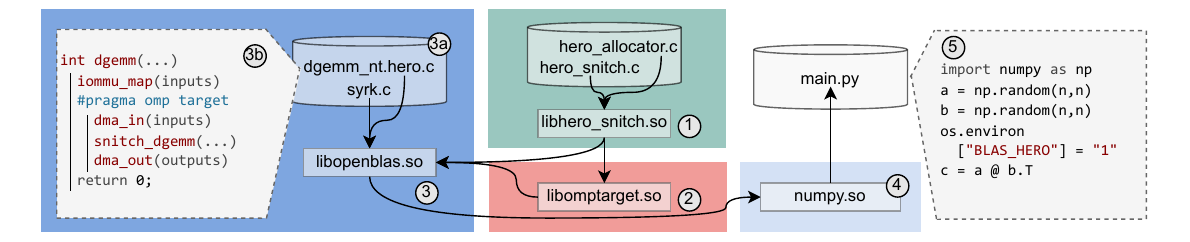}
    \caption{The proposed software architecture. The Hero library \textcircled{1} contains device managements functions. The OpenMP target library \textcircled{2} contains the callbacks for the OpenMP API. The OpenBLAS library \textcircled{3} contains computing kernels for host and/or device. The Numpy \textcircled{4} package is linked against OpenBLAS. Finally, the user application \textcircled{5} imports Numpy.}
    \label{fig:sw}
\end{figure*}

\section{Open-Source Software Stack}

For this study, we introduce the software stack shown in Figure\,\ref{fig:sw}. This stack is built upon previous work for heterogeneous programming.

\subsubsection{HeroSDK:}

Hero\gls{sdk} \cite{herosdk} contains a heterogeneous LLVM ($15$) compiler, host runtimes, and Linux kernel ($6.1.22$) modules to build heterogeneous applications with OpenMP. The Hero library \textcircled{1} contains device management functions and calls to the kernel module (for instance, to map the accelerator's IO\,space). This library contains device-agnostic files such as \textit{hero\_allocator.c}, which manages L2 \gls{spm} and device \gls{dram}, and device-specific files like \textit{hero\_snitch.c} which boots the device. This allows for quick porting to new open-source accelerators. Then, the OpenMP target library \textcircled{2} implements callbacks for the device offload API via the LibHero. As this stack is implemented in C/C++, it may not be easily usable in high-level applications. Thus, in this work, we extend Hero\gls{sdk} with OpenBLAS support.

\subsubsection{Compiling OpenBLAS with HeroSDK:}

We included into OpenBLAS ($0.3.29$) an heterogeneous \texttt{rv64}/\texttt{rv32} implementation of \gls{gemm}. With minor changes to the Makefiles in OpenBLAS, we select kernels to be compiled only for the host like \textit{syrk.c} \textcircled{3a} and kernels to be compiled for the host and accelerator using the Hero\gls{sdk} LLVM compiler. The pseudo-code of the heterogeneous kernel in provided in \textcircled{3b}. The execution starts on the host and the region within the \texttt{\#pragma} is compiled and offloaded to the accelerator. The resulting \textit{libopenblas.so} \textcircled{3b} contains the device functions to be copied to L2 \glsunset{spm} before the first offload.

\subsubsection{Python test application:}
We can link Numpy to OpenBLAS \textcircled{4}, and write Python applications \textcircled{5}.

\section{Results}

\begin{figure}[t]
\centering
\includegraphics[width=0.92\columnwidth]{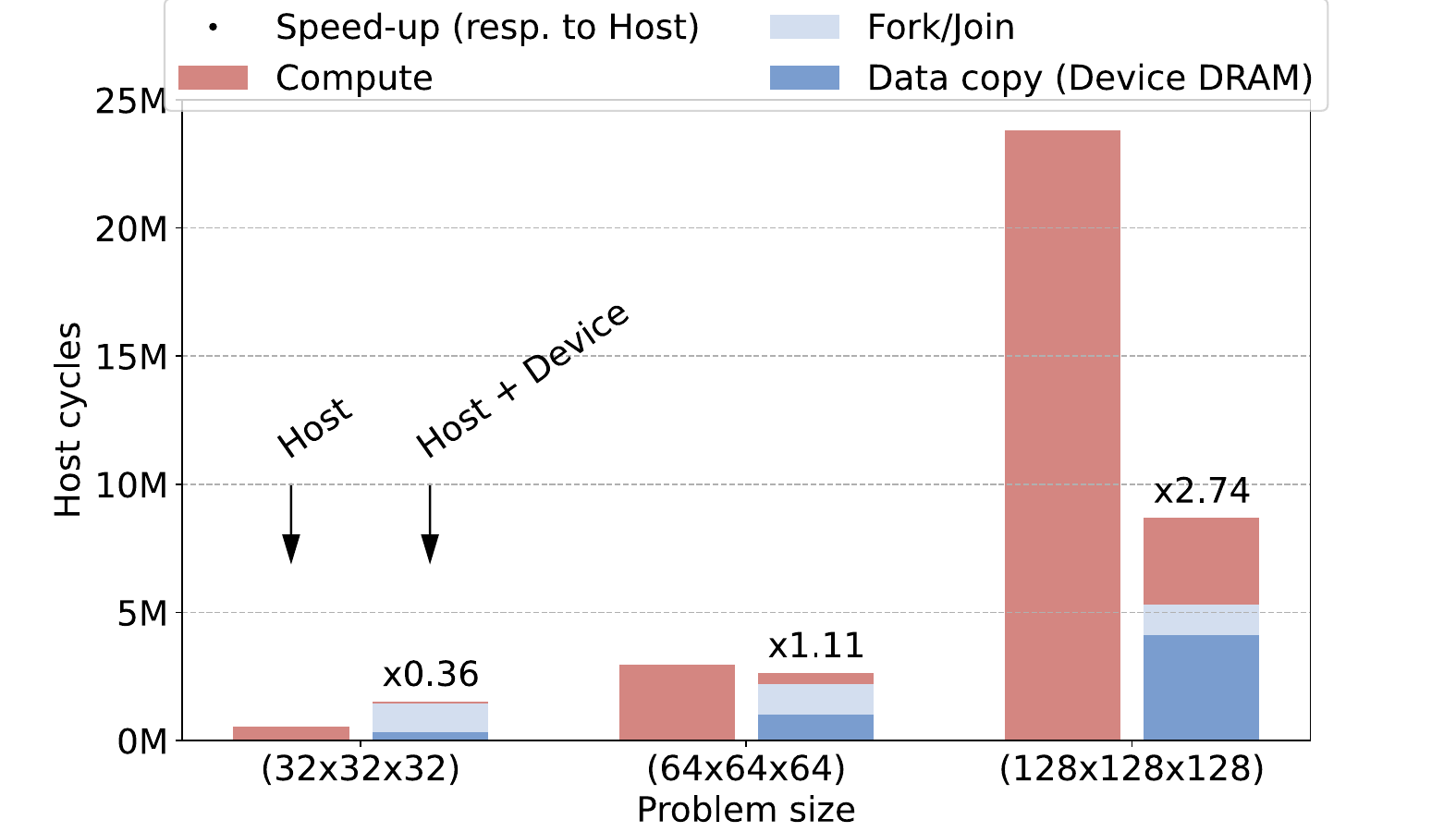}
%\includesvg[width=0.92\columnwidth]{fig/result.svg}
\caption{Execution time (measured from Python) for a $float64$ matrix multiplication with and without offloading.}
\label{fig:result}
\end{figure}

We execute the application \textcircled{5} with and without Hero device offloading and measure the execution time using the Python function \textit{os.time()}. In Figure\,\ref{fig:result}, we show the runtime divided into three regions for different problem sizes. During the \textit{"data copy"} region, the host copies inputs and outputs between the Linux portion and the device portion of the \gls{dram}. During \textit{"fork/join"} the host enters and exits OpenBLAS and OpenMP. In \textit{"compute"} the device \gls{dma} copies local data and processes them in \gls{spm}. One of the key challenges of heterogeneous computing is to keep the \textit{"fork/join"} and \textit{"data copy"} overheads as low as possible to reach interesting speedups. Our solution is $2.71\times$ faster than host-only computation for matrices of size $128$. The \textit{"data copy"} remains the major overhead with $47\%$ of the total runtime. Since the platform features an open-source RISC\nobreakdash-V \gls{iommu}\,\cite{pinto_iommu_2023}, future work will focus on removing this overhead via zero-copy offloading. From a previous study on the same platform, we expect creating IO page table entries for this input size to be $7.5\times$ faster than copying, bringing the total speedup to $4.7\times$. Further improvements can be expected from highly optimized kernels and \gls{simd} operations on lower precision data types.

\section{Discussion}
In this preliminary work, we compile OpenBLAS for an open-source RISC\nobreakdash-V \gls{hesoc}. With minor modifications to the codebase, we extend the existing RISC\nobreakdash-V host implementation with heterogeneous kernels for RISC\nobreakdash-V \glspl{pmca}. We verify the benefits of the accelerator on a simple Python application with Numpy. This allows for easily leveraging heterogeneous RISC\nobreakdash-V SoCs in high-level applications such as ML frameworks.

%----------------------------------------------------------------------------------------
%	 REFERENCES
%----------------------------------------------------------------------------------------

\printbibliography % Output the bibliography

%----------------------------------------------------------------------------------------

\end{document}

%% file: acronyms.tex
\newacronym{abi}{ABI}{application binary interface}
\newacronym{ai}{AI}{artificial intelligence}
\newacronym{api}{API}{application programming interface}
\newacronym{apu}{APU}{accelerated processing unit}
\newacronym{asic}{ASIC}{application-specific integrated circuit}
\newacronym{axi}{AXI}{advanced extensible interface}

\newacronym{blas}{BLAS}{Basic Linear Algebra Subprograms}

\newacronym{cccc}{CCCC}{C and C++ Code Counter}
\newacronym{cnn}{CNN}{convolutional neural network}
\newacronym{cpu}{CPU}{central processing unit}
\newacronym{c2c}{C2C}{chip-to-chip}

\newacronym{dnn}{DNN}{deep neural network}
\newacronym{dma}{DMA}{direct memory access}
\newacronym{dram}{DRAM}{dynamic random access memory}
\newacronym{dsa}{DSA}{domain-specific accelerator}

\newacronym{eoc}{EOC}{end of computation}
\newacronym{epac}{EPAC}{European processor accelerator}

\newacronym{fdt}{FDT}{flattened device tree}
\newacronym{fpga}{FPGA}{field-programmable gate array}
\newacronym{flops}{FLOPs}{floating-point operations per second}
\newacronym{fpu}{FPU}{floating-point unit}

\newacronym{gemm}{GEMM}{GEneral Matrix Multiply}
\newacronym{gpu}{GPU}{graphical processing unit}

\newacronym{hart}{hart}{hardware thread}
\newacronym[plural=heSoCs, longplural=heterogeneous systems-on-hip]{hesoc}{heSoC}{heterogeneous system-on-chip}
\newacronym{hpc}{HPC}{high performance computing}
\newacronym{hls}{HLS}{high level synthesis}

\newacronym{idol}{I\$}{instruction cache}
\newacronym{iommu}{IOMMU}{IO memory management Unit}
\newacronym{iotlb}{IOTLB}{IO translation lookaside buffer}
\newacronym{iova}{IOVA}{IO virtual address}
\newacronym{isa}{ISA}{instruction set architecture}

\newacronym{llc}{LLC}{last-level cache}
\newacronym[longplural={networks-on-chip}]{noc}{NoC}{network-on-chip}
\newacronym{loc}{LOC}{lines of code}
\newacronym{lsu}{LSU}{load store unit}

\newacronym{mac}{MAC}{multiply and accumulate}
\newacronym{mimd}{MIMD}{multiple-instructions-multiple-data}
\newacronym{ml}{ML}{Machine Learning}
\newacronym{mmu}{MMU}{memory management unit}
\newacronym{mpsoc}{MPSoC}{multi-processor SoC}

\newacronym{os}{OS}{operating system}

\newacronym{rab}{RAB}{remapping address block}
\newacronym{rtt}{RTT}{round-trip time}

\newacronym{pci}{PCI}{peripheral component interconnect}
\newacronym{pcie}{PCIe}{peripheral component interconnect express}
\newacronym{pe}{PE}{processing element}
\newacronym{pmca}{PMCA}{programmable manycore accelerator}
\newacronym[plural=PTEs, longplural=page table entries]{pte}{PTE}{page table entry}
\newacronym{ptw}{PTW}{page table walk}

\newacronym{sdk}{SDK}{software development kit}
\newacronym{simd}{SIMD}{single-instruction-multiple-data}
\newacronym{simt}{SIMT}{single-instruction-multiple-threads}
\newacronym{slc}{SLC}{system-level cache}
\newacronym[plural=SoCs, longplural=systems-on-chip]{soc}{SoC}{system-on-chip}
\newacronym[longplural={scratchpad memories}]{spm}{SPM}{scratchpad memory}
\newacronym{spmd}{SPMD}{single program multiple data}
\newacronym{sram}{SRAM}{static random access memory}

\newacronym{tcdm}{TCDM}{tightly-coupled data memory}
\newacronym{tlb}{TLB}{translation lookaside buffer}